\documentclass[twocolumn,amsmath,amssymb,nofootinbib,superscriptaddress,showpacs]{revtex4}

\usepackage{graphicx}
\usepackage{epstopdf}
\usepackage{bm}
\usepackage{color}

\newcommand{\e}{\mathrm{e}}

\newcommand{\prob}{\mathcal{P}}

\newcommand{\lc}{\lambda_c}
\newcommand{\lcNL}{\lambda_c(N,L)}

\newcommand{\kave}{\langle k\rangle}
\newcommand{\ksqave}{\langle k^2\rangle}

\begin{document}

\title{Epidemic spreading in evolving networks}

\author{Yonathan Schwarzkopf}
\affiliation{Department of Physics of Complex Systems,
Weizmann Institute of Science,
Rehovot, Israel 76100}
\affiliation{California Institute of Technology, Pasadena, CA 91125}
\affiliation{Santa Fe Institute, Santa Fe, NM 87501}
\author{Attila R\'akos}
\affiliation{Department of Physics of Complex Systems,
Weizmann Institute of Science,
Rehovot, Israel 76100}
\affiliation{Research Group for Condensed Matter Physics of the Hungarian Academy of Sciences, \\ Budapest University of Technology and Economics, 1111 Budapest, Hungary}
\author{David Mukamel}
\affiliation{Department of Physics of Complex Systems,
Weizmann Institute of Science,
Rehovot, Israel 76100}

\date{\today}

\begin{abstract}
A model for epidemic spreading on rewiring networks is introduced
and analyzed for the case of scale free steady state networks. It is
found that contrary to what one would have naively expected, the
rewiring process typically tends to suppress epidemic spreading. In
particular it is found that as in static networks, rewiring networks
with degree distribution exponent $\gamma >3$ exhibit a threshold in
the infection rate below which epidemics die out in the steady
state. However the threshold is higher in the rewiring case. For
$2<\gamma \leq 3$ no such threshold exists, but for small infection
rate the steady state density of infected nodes (prevalence) is
smaller for rewiring networks.
\end{abstract}

\pacs{89.90.+n,89.75.-k,05.40.-a}

\maketitle

\section{Introduction}

Epidemic spreading can be thought of as occurring on complex
networks where the nodes of the network represent individuals and
the links represent various interactions among those individuals.
For example the spreading of diseases can be thought of as occurring
over the network of human contacts \cite{liljeros-2001-411} and the
spreading of computer viruses as occurring over the internet
\cite{may-2001,newman-2002}. Models of epidemic spreading over
networks have been studied extensively in recent years (for reviews
see \cite{Dorogovtsev08,Barrat08}). Typically, the underlying
network in these models is considered to be static while the state
of the individuals residing on its nodes can change from infected to
non-infected according to some dynamical rules. One is then
interested in studying the evolution of an infected region in time,
the average density of infected nodes in steady state (prevalence)
and the way they are affected by the statistical properties of the
network and the infection rates.

In general, networks can be characterized by the connectivity of their
nodes. The connectivity (degree) $k$ of a node is defined as the
number of links connected to the node. The degree distribution of a
network $\prob(k)$ is defined as the probability of a randomly
chosen node to have a degree $k$. Many networks such as social
networks, the internet and the World Wide Web (WWW) have been found
to be scale free (SF) \cite{barabasi_science_1999,barabasi_1999,www_sf_BA,Albert02},
meaning that the degree distribution follows a power law
\begin{equation}\label{eq_degree_power}
\prob(k) \sim k^{-\gamma}~.
\end{equation}
In the thermodynamic limit one can divide SF networks into two
classes based on the exponent $\gamma$. For $ \gamma
>3 $ the second moment of the degree distribution is finite and as such the system exhibits
finite degree fluctuations. For \mbox{ $2 < \gamma \leq 3$ } the
second moment diverges resulting in infinitely large degree
fluctuations. 
In the present study we only consider networks with a finite degree distribution corresponding to $\gamma>2$.
Interestingly, many real networks have been measured to belong to
the second class having $2 < \gamma \leq 3$ \cite{Albert02}.

Studies of models of epidemic spreading over static networks have
shown that in networks for which $\gamma >3$, the prevalence, $\rho$
vanishes for sufficiently small infection rates $\lambda$. The
prevalence become non-zero only beyond a threshold rate $\lambda_c$.
On the other hand for networks with $2 < \gamma \leq 3$, for which
the second moment of the degree distribution diverges, the
prevalence is non-zero for any infection rate, and no threshold
exists \cite{Satorras01a,Satorras01b,Satorras02b,Boguna03b}. Thus,
epidemics are easier to stop in static networks with $\gamma>3$.

In many cases networks are not static but rather evolve in time, for
example via rewiring processes. Steady states of rewiring networks
have been studied in the past. It has been shown that depending on
the average degree and the rewiring rates, networks may reach an SF
steady state, with an exponent $\gamma$ which can be expressed in
terms of the dynamical rates \cite{Dorogovtsev03,zrp_ref,Angel05}.

In the present paper we consider epidemic spreading over rewiring
networks. On such networks, the disease can spread at a given time
through the links which are present at that time. We find that as in
the static case a non-vanishing threshold value of the infection
rate, $\lambda_c$, exists for $\gamma >3$. Below this threshold the
prevalence (fraction of infected individuals) vanishes while above
it the prevalence is non-zero. For $2 < \gamma \le 3$ no such
threshold exists and the steady state prevalence is non-zero for any
$\lambda
>0$. However, contrary to what one would have naively expected,
epidemic spreading in our model is not necessarily enhanced by the dynamics of
the network. For $\gamma > 3$ the threshold $\lambda_c$ is found to
be larger than that of the corresponding static network. Also, for
$2 < \gamma \le 3$ the prevalence at small $\lambda$ is found to be
smaller than that of the corresponding static network.

The paper is organized as follows: in section \ref{review} we review
known results on epidemic spreading in static networks and on
networks with rewiring dynamics. In section \ref{epidemic} we study
epidemic spreading on evolving networks using mean field
calculations and numerical simulations. Our results are summarized
in section \ref{conclusions}.

\section{Review of known results}
\label{review}

\subsection{Epidemic spreading in static networks}

A number of models of disease spreading have been introduced and
studied in the past. In the present work we use the Susceptible
Infected Susceptible (SIS) model
\cite{Satorras01a,Satorras01b,Satorras02a,Albert02,Dorogovtsev03,Dorogovtsev08,Barrat08}.
In this model a healthy individual, with respect to the disease, may
be infected through interaction with diseased individuals. Meaning,
that a susceptible node may be infected through a link connecting it
to an infected node, which we will refer to as his neighbor. Once an
individual is infected he may become susceptible again by being
spontaneously cured from the disease. The curing process does not
immune the individual and it can be reinfected.

The continuous time dynamics of an epidemic in the SIS model is defined by two
stochastic processes using two parameters:
\begin{description}
\item[] $\lambda$- Infection rate%
 \item[]$\delta$ -  Rate of recovery
 \end{description}
An infected node is spontaneously cured
with a rate $\delta$ which we choose to be equal to 1 by adjusting the time scale.
On the other hand a susceptible node gets infected with rate $\lambda$ from
each of its infected neighbors. Thus, the rate a node
is infected depends linearly on the number of infected neighbors.
This model of infection is different from the model explored in
\cite{Dorogovtsev03,Satorras01a,Satorras01b,Satorras02a} where the
infection rate is independent of the number of infected neighbors.
However, both models behave similarly  near the threshold for an
endemic state and we expect our conclusions to hold for both models.

The problem is addressed using a mean field (MF) approach and
numerical simulations. The MF approach neglects correlation in
infection between nodes in the sense that for any pair of nodes
$i,j$ we have \mbox{$\langle\eta_i\eta_j\rangle=\langle\eta_i\rangle\langle\eta_j\rangle$} where
$\eta=0,1$ is a parameter indicating whether a node is susceptible
or infected, respectively. As an order parameter we use the
prevalence of the disease, the density of infected nodes in the
network, defined as $\rho\equiv N_\text{infected}/N$. Hence, our
problem is reduced to
  a contact equation for the order parameter $\rho$.
  Since we are interested in formulating the problem for any degree distribution, as was
previously done in \cite{Satorras01a,Satorras01b}, we
shall distinguish between nodes of different degree by defining $\rho_k$
as the fraction of diseased nodes of degree $k$. The total
prevalence is thus given by
\begin{equation}
\rho=\sum_{k=0}^{\infty}\rho_k \prob(k)
\end{equation}

The MF contact equation has the following form
\begin{equation}\label{eq_contact}
\frac{\partial \rho_k}{\partial t}=-\rho_k+\lambda k
\ell(1-\rho_k)
\end{equation}
With $\ell$ being the density of ``infected links'' defined as
\begin{equation}\label{eq_ell}
\ell=\sum_{k=0}^{\infty}\frac{k \prob(k)}{\kave} \rho_k
\end{equation}
Note that
\begin{equation}
 \prob^*(k)=\frac{k\prob(k)}{\kave}
\end{equation}
is the degree distribution of a randomly chosen neighbor of nodes.
Thus \eqref{eq_ell} gives the probability that a randomly chosen end
of a randomly chosen link is infected. In the steady state, a non-
vanishing solution for the prevalence is possible only for infection
rates greater than (see \cite{Satorras01a,Satorras01b})
\begin{equation}\label{eq_lc}
\lc=\frac{\kave}{\ksqave}.
\end{equation}
For infection rates above the threshold,
\mbox{$\lambda>\lc$}, a finite fraction of the nodes is infected
while for \mbox{$\lambda \leq \lc$} the disease dies out\footnote{
In the thermodynamic limit this is a transition to an absorbing
state, but for finite size systems the only true steady state is one
with zero prevalence \cite{Satorras02b}. As a result, in finite
networks there is no true threshold but a crossover infection rate
which can be calculated for a quasi-stationary state.
 }.

For Erd\H{o}s-R\'{e}nyi (ER) networks, which obey a Poisson degree
distribution \cite{erdos-1960}, the threshold can be rewritten in
the form \mbox{$\lc=1/(\kave+1)$}. Moreover,  for SF networks with
$2<\gamma\leq 3$ the second moment diverges and as a result the
threshold vanishes. As a consequence such a system will always reach
an endemic steady state for any non zero infection rate $\lambda
>0$.


\subsection{networks under rewiring dynamics}
\begin{figure}
\includegraphics[width=4cm]{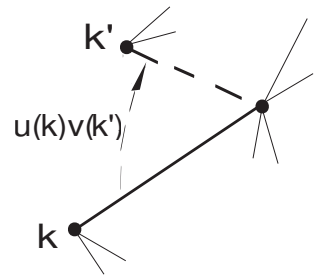}
\caption{Rewiring of a link from a node with degree $k$ to a node
with degree $k'$ with a rate $u(k)v(k')$.}\label{fig_rewire}
\end{figure}

During rewiring dynamics of a network the number of nodes and the number of links are
unchanged but the links are stochastically detached from one node
and  reattached to another. In our model the process of rewiring a randomly
chosen end of a link from a node with degree
$k$ to a node with degree $k'$ occurs
with rate $u(k)v(k')$. A schematic representation of the process
is given in Fig.~\ref{fig_rewire}.
 These rates determine the
steady state degree distribution of the network through the
relation
\begin{equation}\label{eq_dorogorov_prob}
\prob(k)=\frac{\prod_{k'=0}^{k-1}
v(k')}{\langle v \rangle^k}\frac{\langle u\rangle^k}{\prod_{k'=1}^k u(k')}\prob(0)
\end{equation}
which can be derived from the master equation for the
node degree distribution \cite{Dorogovtsev03}.

Under such rewiring dynamics, the resulting networks are
uncorrelated in the sense that the joint probability $\pi(k,k')$ that the
ends of a randomly chosen link are nodes of degree $k$ and
$k'$ factorizes to
\begin{equation}
 \pi(k,k')=\prob^*(k)\prob^*(k')
\end{equation}

By choosing the proper attachment and detachment rates one can
create an evolving network with a constant size and any desired
degree distribution. One such choice of rewiring  yields an evolving
ER type network. This is achieved by choosing a link at random and
rewiring one of it's randomly chosen ends to a randomly chosen node.
This rewiring scheme has a constant attachment rate and a linearly
preferential detachment rate. One can easily verify by using
(\ref{eq_dorogorov_prob}) that the choice
\begin{eqnarray}\label{eq_rates_ER}
v(k)&=& \frac1N \nonumber \\
u(k)&=&k
\end{eqnarray}
indeed yields a
Poisson degree distribution.

Through the use of such rewiring dynamics we can create an
uncorrelated SF network with any desired exponent in the power law
distribution. In what follows we work with rewiring dynamics similar
to that of zero-range processes (ZRP) \cite{Angel05,zrp_ref}, where the
rewiring rate does not depend on the destination site, i.e.,
$v(k)=1/N$. As a further specification we consider detachment rates of
the form
\begin{equation}
\label{eq_hopping_rate_SF}
u(k) = 1+\frac{b}{k}
\end{equation}
with $b$ as a parameter of the dynamics. In this case, for a
specific choice of the average number of links (given by
$\kave=1/(b-2)$) the underlying zero-range process exhibits critical
behavior in which the steady state degree distribution is a power
law $\prob(k) \sim k^{-b}$ at large $k$. At lower average link
numbers the steady state distribution decays exponentially with $k$
while at larger averages a hub becomes present which is linked to a
finite fraction of the nodes in the network \cite{Angel05,zrp_ref}.

In order to be able to control the critical value of $\kave$ for a
given value of $b$ one can make a slight modification in the
dynamics by considering
\begin{equation}\label{eq_hopping_rate_n0}
u(k) =
\begin{cases}
1+\frac{b}{k_0} & 0<k\leq k_0\\
1+\frac{b}{k} & k>k_0
\end{cases}
\end{equation}
Due to the same asymptotic behavior of \eqref{eq_hopping_rate_SF}
and \eqref{eq_hopping_rate_n0} this modification does not change the
power law tail of the stationary degree distribution. In this case the critical value of $\kave$ can be obtained numerically \cite{zrp_ref}. Note that for rewiring rates of the form \eqref{eq_hopping_rate_SF} and \eqref{eq_hopping_rate_n0}, $\langle u \rangle=1$ at criticality \cite{zrp_ref}.

It is important to note that the dynamics, as defined, allows for
multiple link between two nodes (melons) and links that connect a
node with itself (tadpoles). By not allowing for melons and tadpoles
we are introducing an effective preferential attachment rate
$v(k)=1-k/N$ as opposed to a constant rate as given in
(\ref{eq_hopping_rate_SF}). This rate takes into account the fact that the neighbors of a node
of degree $k$ are not available as target nodes for the rewired link.
The preferential attachment rate means
that a highly connected node has a lower rate of attachment than a
node with a lower connectivity and induces disassortative, or negative
correlations. It can be
shown using (\ref{eq_dorogorov_prob}) that this attachment rate
imposes a Gaussian cutoff on the degree
distribution of the form
\begin{equation}\label{eq_noloops_cutoff}
\prob'(k)=\e^{-\frac{k^2}{2N}}\prob(k)
\end{equation}
where $\prob(k)$ is the degree distribution for similar dynamics
which allows for tadpoles and melons. For ER type networks and for SF networks with
$\gamma >3$ the fraction of melons and tadpoles vanishes in the
thermodynamic limit \cite{Boguna04}. However, for SF networks with $\gamma \leq 3$
the number of melons and tadpoles diverges and cannot be neglected.
Since the infection process was taken to depend linearly on the
number of infected links, the problem could be restated by choosing
networks with weighted links and an infection process which depends
linearly on the weight of the link. 
\section{Epidemic spreading on evolving networks}
\label{epidemic}

Our aim is to consider a model of epidemic spreading on a network
which is changing in time. As a consequence, a given node is no
longer connected to a static set of neighbors but to a dynamic one,
and the degree $k$ of the node also fluctuates.
Previous work on epidemic spreading  on evolving networks \cite{Gross06,Volz07,Fefferman07,shaw2008,shaw2009} concentrated mainly on rewiring dynamics resulting from the adaptation of the network to the disease. In these models the rewiring dynamics depend on the state of the nodes, i.e. the infection process.
 We consider models
where the rewiring dynamics is independent of the infection process.
To be more specific, we consider a ZRP-like rewiring dynamics for
the network with rates of the form \eqref{eq_hopping_rate_n0} and $v(k)=1/N$.
For a specific value of $\langle k\rangle$ this results in a power-law degree distribution for the steady state
of the network (which depends on $k_0$) with ${\cal P}(k)\sim k^{-b} $. In addition, we introduce a parameter $\nu$,
which describes the overall timescale of the rewiring process as compared to that of
the infection process. The rewiring rate from a node of degree $k$ then becomes $\nu u(k)$. For
$\nu=0$ the model reduces to epidemic spreading on a static network,
whereas for $\nu\to\infty$, due to the fast mixing we expect a mean-field like behavior for
the infection process, where neighbors change very rapidly. We note here that we always consider the case
where the network is in a stationary state with respect to the
rewiring dynamics, which requires a diverging equilibration time for
$\nu\to 0$.

\subsection{Mean-field results}
In order to account for the rewiring dynamics \eqref{eq_contact} has
to be modified as follows:
\begin{multline} \label{eq_contact_rew}
\frac{\partial \rho_k}{\partial t}=-\rho_k+\lambda k\ell(1-\rho_k)- \nu \rho_k(u_k+\langle u\rangle) \\
+\nu\left[ \rho_{k+1}\frac{\prob(k+1)}{\prob(k)}u_{k+1} + \rho_{k-1}\frac{\prob(k-1)}{\prob(k)}\langle u\rangle \right]
\end{multline}
By multiplying \eqref{eq_contact_rew} with $\prob(k)$ and summing up for all $k$ one obtains
\begin{equation}\label{eq_rho_ss}
 \rho=\lambda\langle k \rangle \ell (1-\ell)
\end{equation}
in the stationary state.
For infinitesimal $\rho$ and $\ell$ (at the threshold) this reduces to
\begin{equation} \label{eq_rho_lambda}
 \rho=\lambda_c\langle k \rangle \ell.
\end{equation}
Note that one can rewrite definition \eqref{eq_ell} of $\ell$ as
\begin{equation}\label{equiv}
 \ell\equiv\rho{\kave_\text{inf}}/{\kave},
\end{equation}
where $\langle\cdot\rangle_\mathrm{inf}$ denotes an average in the
ensemble of {\em infected} nodes.
We define the average of a quantity
$x$ in the ensemble of {\em infected} nodes as
\begin{equation}
\langle x\rangle_\mathrm{inf}=\rho^{-1}\,\sum_k x \rho_k\prob(k).
\end{equation} 
Using this, \eqref{eq_rho_lambda} takes the following simple form
\begin{equation}\label{eq_rho_lambda2}
 {\lambda_c}^{-1}=\langle k \rangle_\mathrm{inf}.
\end{equation}

On the other hand, multiplying \eqref{eq_contact_rew} by $k\prob(k)$
and summing up over all $k$ one obtains using (\ref{eq_rho_lambda}) for the steady state
\begin{equation} \label{eq_lambdac_rew}
{\lambda_c}^{-1} = \frac{\langle k^2 \rangle}{\langle k \rangle}
+\nu \left( \langle u \rangle - \langle u
\rangle_\mathrm{inf}\right).
\end{equation}

Whether or not there exists a non-zero infection rate threshold can
be easily deduced from this equation. For the rewiring rates
(\ref{eq_hopping_rate_n0}) both $\langle u \rangle$ and $\langle u
\rangle_\mathrm{inf}$ are finite. Thus there exists a finite
positive threshold as long as $\langle k^2 \rangle$ is finite,
namely for networks with $\gamma > 3$. In this case the rewiring
rate $\nu$ affects the threshold quite strongly. On the other hand
for networks with $2< \gamma \leq 3$, $\langle k^2 \rangle$ diverges
and $\lambda_c$ vanishes.

It is obvious that for $\nu\to0$ (\ref{eq_lambdac_rew}) reduces to
$\lambda_c=\langle k \rangle/\langle k^2 \rangle$ as discussed
before. In the other extreme case, when $\nu\to\infty$, due to the
fast rewiring we expect that the degree distribution of infected and
non-infected nodes become identical. This would imply that
$\rho_k=\rho$ for all $k$ and $\langle k
\rangle_\mathrm{inf}=\langle k \rangle$. Therefore, based on
(\ref{eq_rho_lambda2}), one has $\lambda_c=\langle k \rangle^{-1}$.
Note also that in this infinite rewiring limit the rhs of
(\ref{eq_lambdac_rew}) is nontrivial, since $\nu\to\infty$, while
$\langle u \rangle - \langle u \rangle_\mathrm{inf}$ is expected to
vanish.

It is important to note that whereas for small values of $\nu$ the
MF approximation, that we use throughout this section, is not
necessarily valid. However in the $\nu\to\infty$ limit the fast
rewiring ruins all the correlations in the system, and the MF
approximation is expected to become asymptotically exact.

As shown by (\ref{eq_rho_lambda2}) the threshold is determined by
the degree distribution of infected nodes {\em at the transition point}.
In the following we attempt to get a deeper understanding of how
this distribution changes with the rewiring rate $\nu$. For this
reason we define
\begin{equation}\label{ansatz}
 r_k=\lim_{\lambda\searrow\lambda_c}\frac{\rho_k}{\rho},
\end{equation}
and assume that this quantity is finite for all $k$.
This implies the following
normalization for $r_k$:
\begin{equation}
 \sum_{k=0}^\infty \prob(k) r_k= 1.
\end{equation}
It is easy to see that
\begin{equation}\label{tildeP}
 \prob(k) r_k = \tilde{\prob}(k)
\end{equation}
is the degree distribution of infected nodes close to the threshold.

Equation (\ref{eq_contact_rew}) together with the steady state relation
\begin{equation}\label{rholambda}
 \frac{\prob(k)}{\prob(k+1)}=\frac{u_{k+1}}{\langle u\rangle}
\end{equation}
implies
\begin{multline}\label{eq_contact_mod}
 0=-\rho_k+\lambda k\ell(1-\rho_k) \\ +\nu\left[\langle u\rangle \rho_{k+1} +u(k)\rho_{k-1}-(\langle u \rangle + u(k))\rho_k\right]
\end{multline}
for the steady state. Inserting \eqref{ansatz} into
\eqref{eq_contact_mod} and using \eqref{rholambda} and \eqref{eq_rho_lambda} one obtains the
following set of equations for $r_k$ at the transition point:
\begin{equation}\label{eq_rk}
0=\frac{k}{\langle k \rangle} -r_k + \nu\left[ r_{k+1} \langle u \rangle +r_{k-1} u_k -r_k\left( u_k + \langle u \rangle \right) \right].
\end{equation}

After solving the above set of equations for $r_k$ one can determine
$\lambda_c$ from \eqref{eq_rho_lambda2} and \eqref{ansatz} as
\begin{equation}\label{num_eval_lambdac}
 \lambda_c^{-1}=\sum_k k r_k\prob(k)
\end{equation}

One can immediately see that in the $\nu\to\infty$ limit the
solution of (\ref{eq_rk}) is $r_k=1$. This implies
$\tilde{\cal P}={\cal P}$ which results in the already noted
limiting behavior with $\lambda_c=1/\kave$. On the other hand, in
the case of a static network, where  $\nu=0$,  $\rho_k$ is
proportional to $k$ near the threshold, implying $r_k=k/\kave$ and
$\tilde\prob=\prob^*$, which results in $\lambda_c=\langle k
\rangle/\langle k^2 \rangle$.

\begin{figure}
\includegraphics[width=7.5cm]{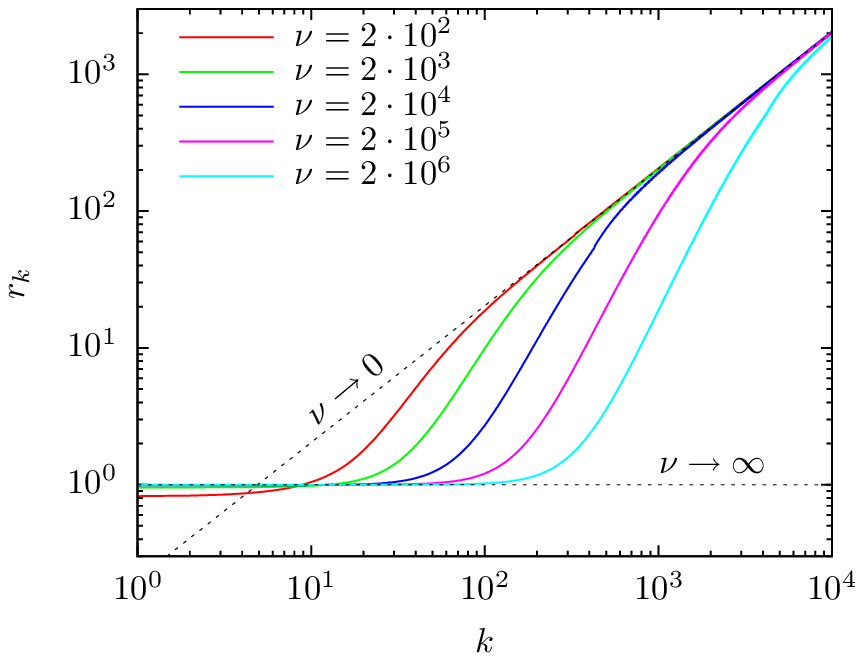}

\vspace*{0.3cm}
\includegraphics[width=7.5cm]{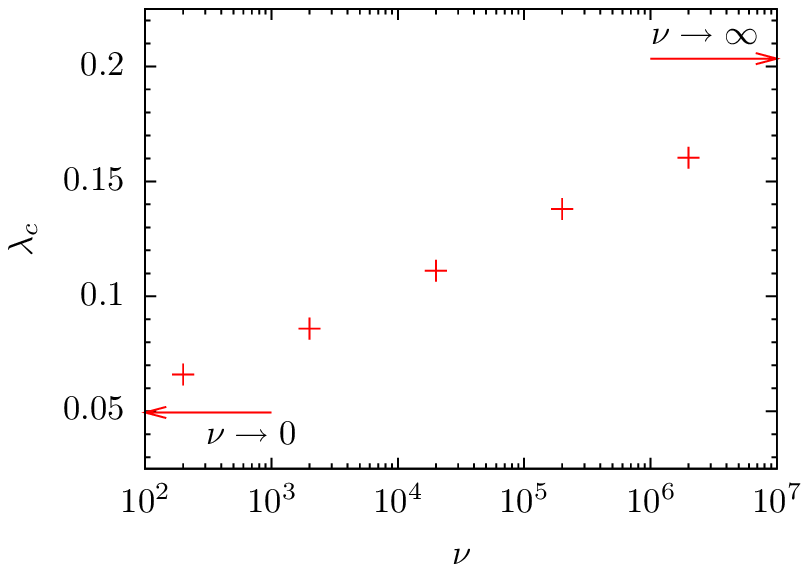}
\caption{\label{fig2} (Top) $r_k$ is plotted against $k$ for various
rewiring rates. The set of equations (\ref{eq_rk}) is solved
numerically with $b=3.5$ and $k_0=15$ for the critical case, which corresponds to $\langle u \rangle=1$, $\langle k\rangle =4.917454$, and $\langle k^2\rangle=99.39410$.
(Bottom) The threshold $\lambda_c$ as a function of $\nu$. It was
calculated numerically using equation (\ref{num_eval_lambdac}) with
the same parameters.}
\end{figure}

Numerical solutions of \eqref{eq_rk} for intermediate values of
$\nu$ are shown in Fig.~\ref{fig2}. One can see that for a large but
finite $\nu$ there are two crossover values of $k$. For $k$ below
some value $k_1$ one has $r_k=1$ and $\tilde\prob=\prob$ (infinite
rewiring), whereas for $k$ larger than some other value $k_2$, one
has $r_k=k/\kave$ and $\tilde\prob=\prob^*$ (static). Between $k_1$
and $k_2$ we find an intermediate regime, which connects the two
extreme cases. The crossover values $k_1$ and $k_2$ increase with
increasing $\nu$.

Using these numerical solutions for $r_k$ we calculated numerically
the effect of rewiring on the threshold $\lambda_c$ by using
\eqref{num_eval_lambdac}. Results are shown in Fig.~\ref{fig2}. One
can clearly see that as the rewiring rate increases the threshold
increases.

\subsection{Simulations}
As discussed above, for SF networks with $2<\gamma \leq 3$ there is
no threshold in the infection rate, and the prevalence is non-zero
for any $\lambda>0$. For SF networks with $\gamma>3$ a threshold exists such that for 
an infection rate below $\lambda<\lambda_c$ the prevalence is zero. 
The prevalence corresponding to such networks
is studied in this section using numerical simulations of finite networks. 

\begin{figure}
\includegraphics[width=8cm]{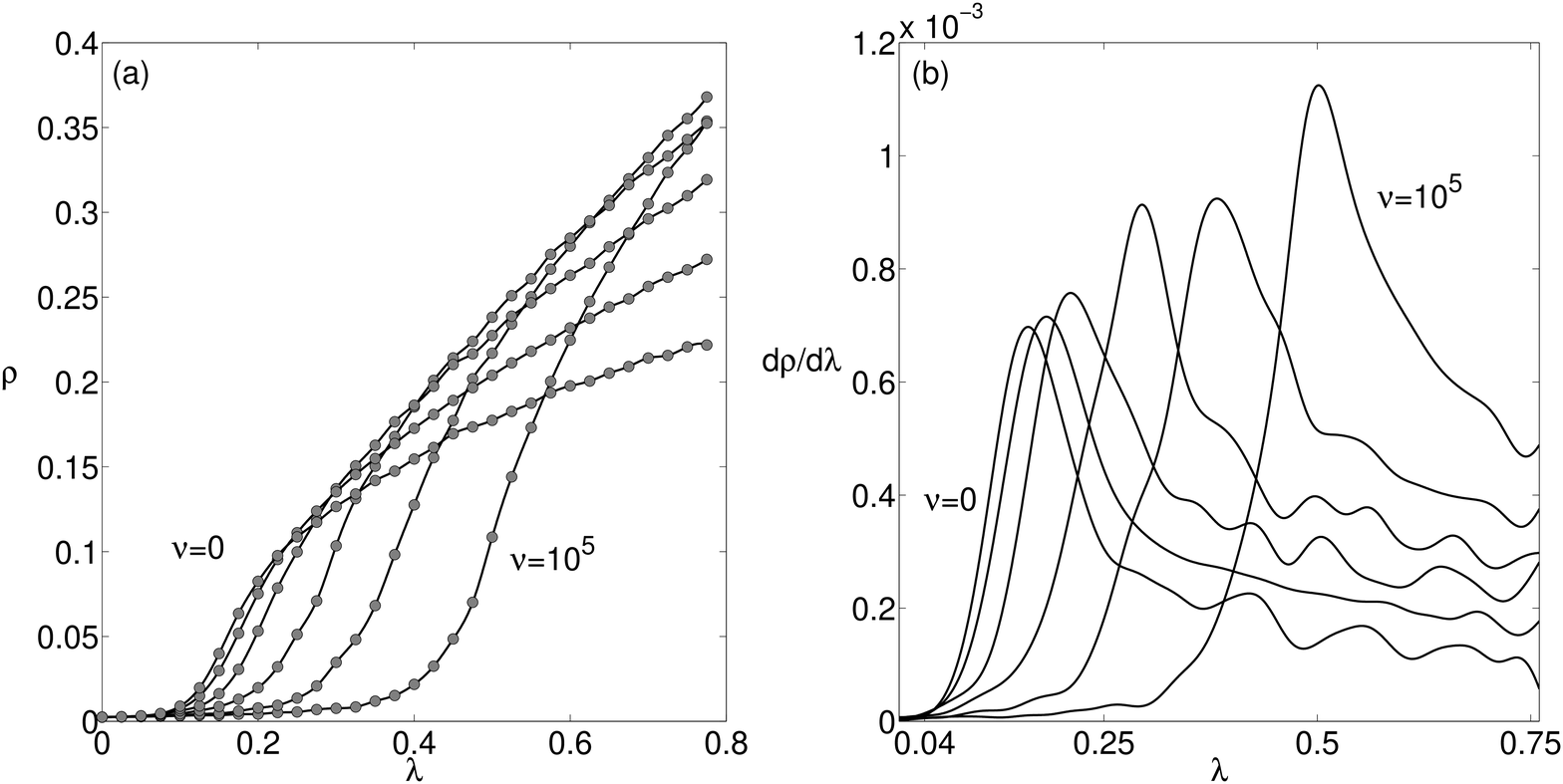}
\caption{ The prevalence $\rho$ (a) and the derivative of the
prevalence with respect to the infection rate $\mathrm{d}\rho
/\mathrm{d} \lambda$ (b) are plotted as a function of the infection
rate for a network of size $N=200$, $\kave=2$ and rewiring dynamics (\ref{eq_hopping_rate_n0})
 with $b=2.5$ and $k_0=1$ for which $\langle k^2\rangle\approx58$.
Curves for rewiring rates \mbox{$\nu=0,10,10^2,10^3,10^4,10^5$} are
represented by full lines from left to right respectively. For
clarity, the derivative was calculated using smoothed interpolated
data of the prevalence.}\label{fig_comp_deriv_to_norm}
\end{figure}

The networks were constructed using two sets of parameter values for 
the rewiring dynamics (\ref{eq_hopping_rate_n0}).
 As an example of a network with $2<\gamma \leq 3$ we use 
the dynamics with
 $v(k)=1/N$, $u(k)=1+b/k$ ($k_0=1$) and $b=2.5$ . As described
previously, $\kave=1/(b-2)=2$ corresponds to the critical average
value of the underlying zero-range process for which the steady
state degree distribution is a power law with $\gamma=2.5$. 
The resulting prevalence
as a function of the infection rate is plotted in
Fig.~\ref{fig_comp_deriv_to_norm}(a) for various rewiring rates.

\begin{figure}
\includegraphics[width=8cm]{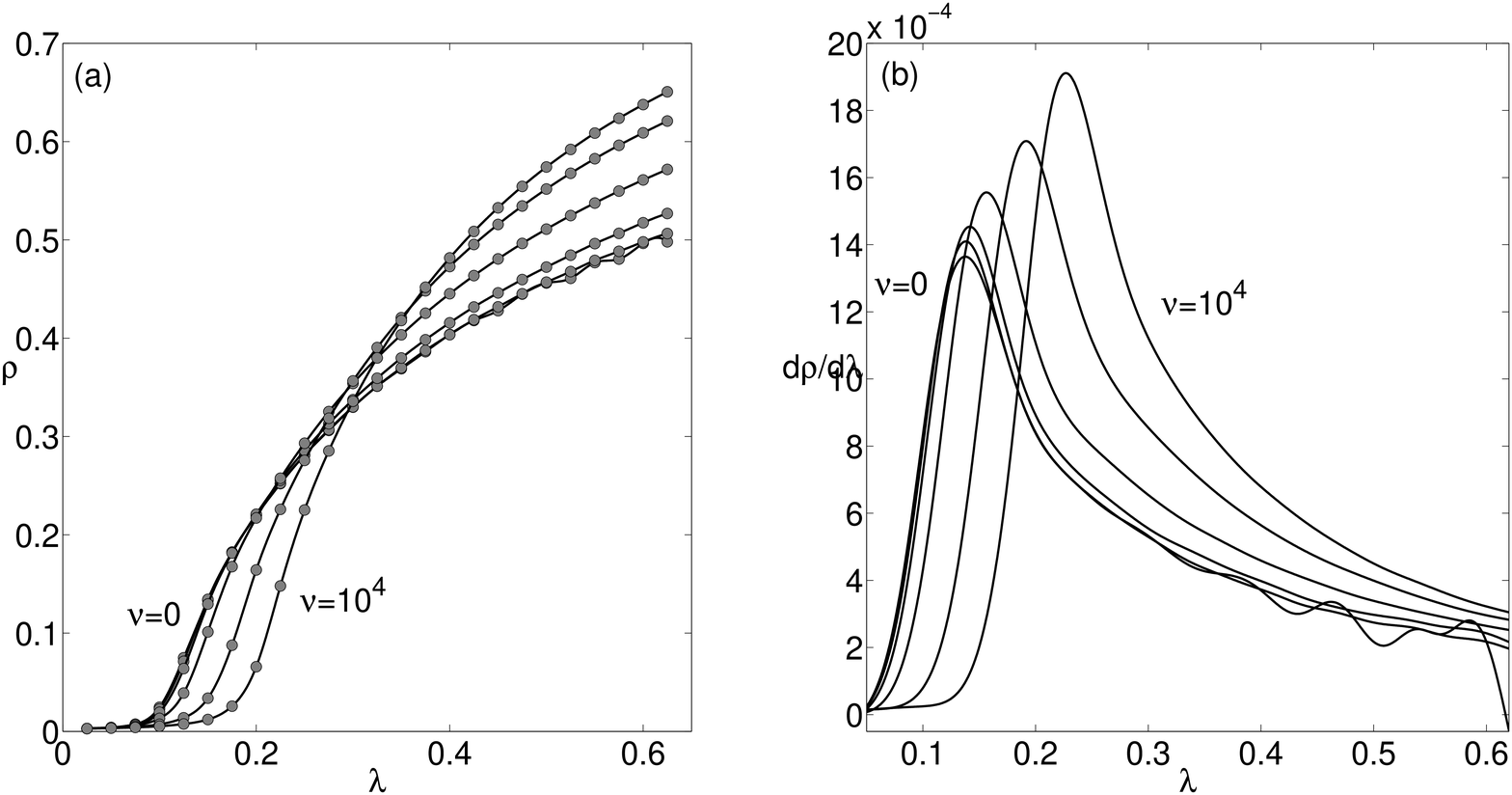}
\caption{ The prevalence $\rho$ (a) and the derivative of the
prevalence with respect to the infection rate $\mathrm{d}\rho
/\mathrm{d} \lambda$ (b) are plotted as a function of the infection
rate for a network of size $N=200$, $\kave=4.9$ and rewiring dynamics
 (\ref{eq_hopping_rate_n0}) with $b=3.5$ and $k_0=15$ for which $\langle k^2\rangle\approx73$.
Curves for rewiring rates \mbox{$\nu=0,10,10^2,10^3$} are
represented by full lines from left to right respectively. For
clarity, the derivative was calculated using smoothed interpolated
data of the prevalence.}\label{fig_comp_deriv_to_norm2}
\end{figure}

As an example of a network with $\gamma > 3$ we used the dynamics
(\ref{eq_hopping_rate_n0}) with $\kave\approx4.917$, $b=3.5$ and $k_0=15$, 
corresponding to the critical average 
value of the underlying zero-range process for which the steady
state degree distribution is a power law with $\gamma=3.5$. 
The resulting prevalence
as a function of the infection rate is plotted in
Fig.~\ref{fig_comp_deriv_to_norm2}(a) for various rewiring rates.

For a network of finite size, there is no true threshold but a
crossover infection rate which is obtained for a quasi-stationary
state. In simulations one can identify this crossover value $\lambda_c(N,L)$ 
as the point where the (numerically obtained) derivative $\mathrm{d}\rho/\mathrm{d}\lambda$ 
takes its maximum. With such a definition $\lambda_c(N,L)\to\lambda_c$ in the thermodynamic limit. 
One can see in Fig.~\ref{fig_comp_deriv_to_norm}(b), corresponding to $2<\gamma<3$,
that as the rewiring rate increases $\lcNL$ increases from
approximately $\kave/\ksqave \approx 0.04$ towards $1/\kave=0.5$.
Similarly, one can see in Fig.~\ref{fig_comp_deriv_to_norm2}(b), corresponding to $\gamma>3$, 
that as the rewiring rate increases $\lcNL$ increases from
approximately $\kave/\ksqave \approx 0.07$ towards $1/\kave=0.2$.

In the simulations a very weak external source of infection was
introduced in order to prevent the system from fluctuating into the
absorbing state. There are several other methods of simulating an
absorbing phase transition and computing from it the value of the
threshold which are reviewed in \cite{qs_ref}. 

\begin{figure}
\includegraphics[width=7cm]{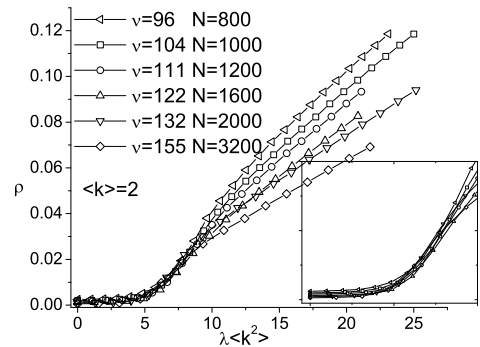}
\caption{The prevalence as a function of the scaled infection rate
$\lc \ksqave$  for a network of different sizes $N$, $\kave=2$ and rewiring dynamics (\ref{eq_hopping_rate_n0})
 with $b=2.5$ and $k_0=1$ plotted for different rewiring rates where for
each network of size $N$ the rewiring rate is $\nu= \ksqave$. The
behavior near the threshold is replotted on a finer scale and is
given in the inset.}\label{fig_data_collapse}
\end{figure}

Note that for the considered value of $b=2.5$, $\ksqave$ diverges in
the thermodynamic limit, therefore, based on the MF results, the
threshold is expected to vanish for $\nu=0$. However, for a finite
system one expects a finite crossover value for $\lambda$, which is
of order $1/\ksqave$. On the other hand, for $\nu\gg\ksqave$ the
crossover value should increase up to $1/\kave$. It is interesting to examine
how the crossover $\lambda$ changes if the rewiring rate is of order
$\ksqave$. To this end we performed simulations with
$\nu\sim\ksqave$. We found that the threshold scales as $\lc^{-1}
\sim \ksqave$. The corresponding data collapse is presented in
Fig.~\ref{fig_data_collapse} where the prevalence is plotted as a
function of a scaled infection rate $\lc\ksqave$ for networks of
different sizes with a rewiring rate equal to the second moment of
the degree distribution $\nu=\ksqave$.

\begin{figure}
\includegraphics[width=9cm]{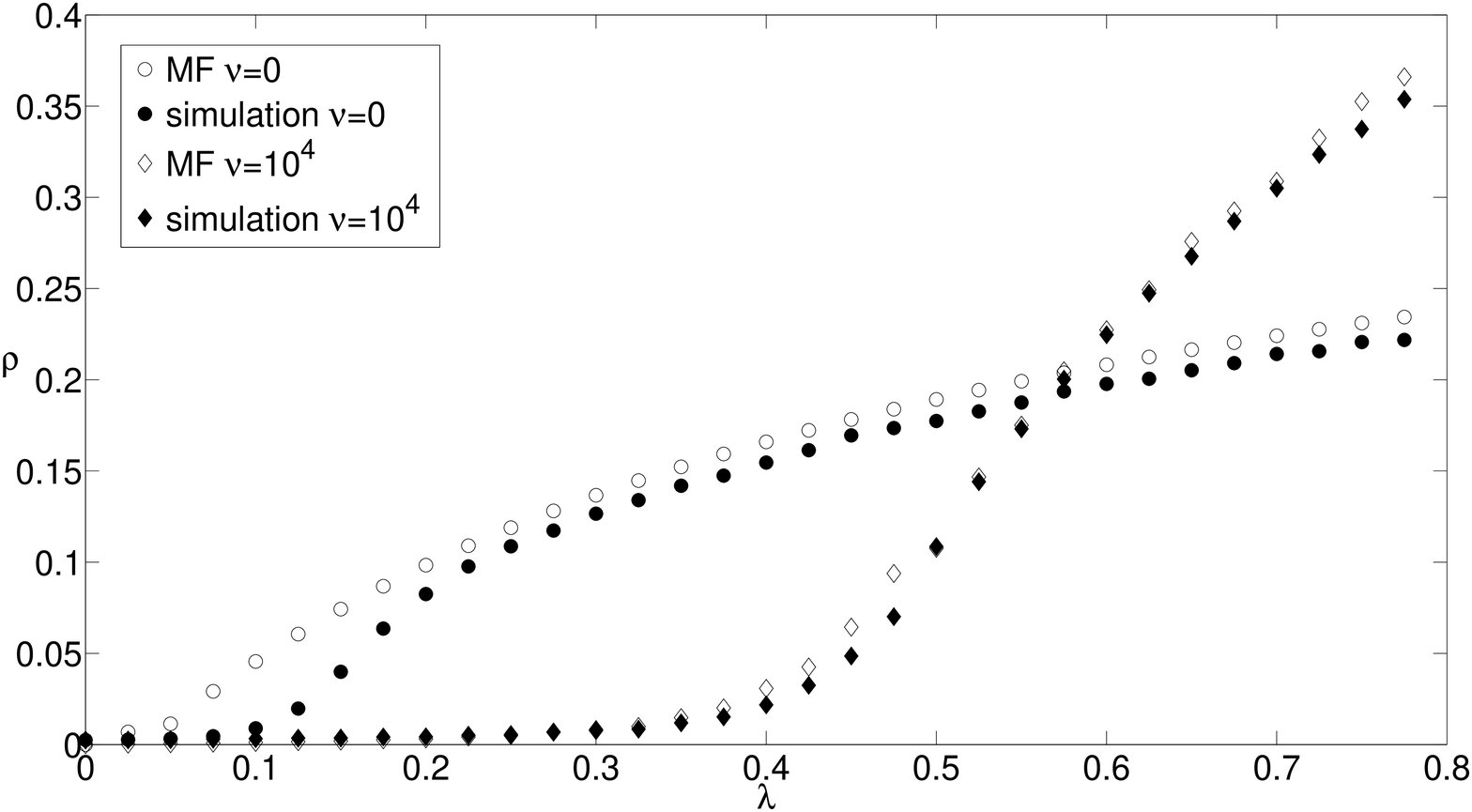}
\caption{The prevalence as a function of the infection rate for a
network of size $N=200$, $\kave=2$ 
and rewiring dynamics (\ref{eq_hopping_rate_n0})
 with $b=2.5$ and $k_0=1$. The
simulation results for $\nu=0$ and $\nu=10^4$ are compared with the numerical
solution of (\ref{eq_rho_ss}). }
 \label{fig:MFb2_5}
\end{figure}

\begin{figure}
\includegraphics[width=9cm]{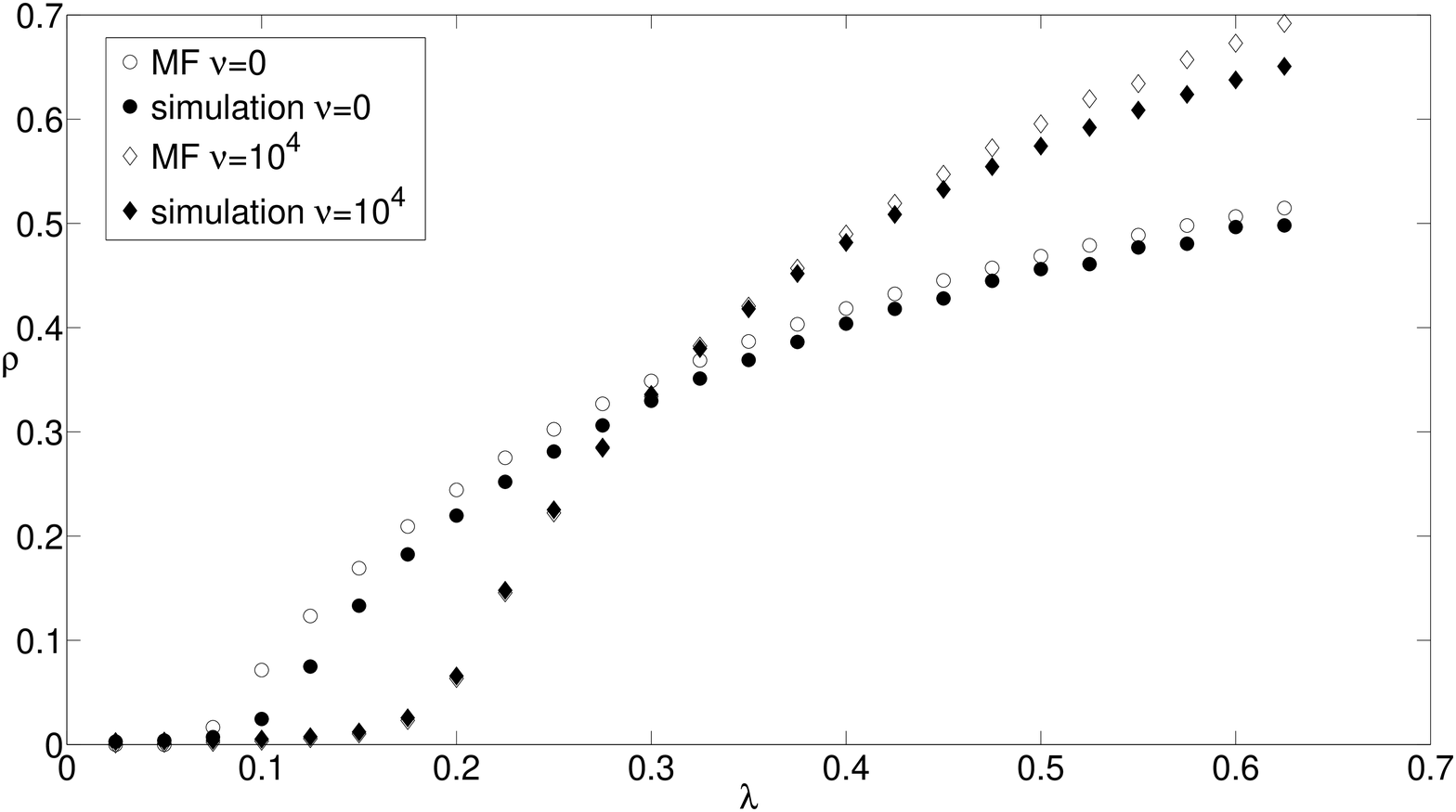}
\caption{The prevalence as a function of the infection rate for a
network of size $N=200$, $\kave\approx4.917$ 
and rewiring dynamics (\ref{eq_hopping_rate_n0})
 with $b=3.5$ and $k_0=15$. The
simulation results for $\nu=0$ and $\nu=10^4$ are compared with the numerical
solution of (\ref{eq_rho_ss}). }
 \label{fig:MFb3_5}
\end{figure}

In Fig.~\ref{fig:MFb2_5} and
Fig.~\ref{fig:MFb3_5} the simulation results are
compared to the numerical solution of (\ref{eq_rho_ss}) for $b=2.5$ and $b=3.5$ for both a static network ($\nu=0$) and for $\nu=10^4$.
The numerical calculation of the MF contact equation was carried out by solving (\ref{eq_contact_rew})
for each infection rate.  The degree distribution used in the
calculation was taken from the simulation results. 
For both the $b=2.5$ and $b=3.5$  cases the MF solution for $\nu=10^4$ agrees quite well with the simulation results which supports our general argument that as the rewiring rate increases, compared to the infection process,  both the degree of a node and fluctuations average out such that the MF approximation better describes the process.

\section{Conclusions}
\label{conclusions}

The effect of network dynamics on epidemic spreading has been
studied using mean field analysis and numerical simulations. In
particular we considered epidemic spreading over SF networks with
rewiring dynamics.

We have shown that the introduction of rewiring affects the
threshold for an endemic state of a network. This is a surprising
result that an evolving network is fitter with respect to disease
the faster it is rewired. This result is general to any network with
a general degree distribution.

One can understand this counter intuitive result by associating
the second moment of the degree distribution with the
heterogeneity of a network. The more heterogeneous is a network
the larger is the fraction of highly connected nodes which mediate
the infection process. The introduction of rewiring effectively
averages out the heterogeneity and creates an effective
homogeneous network, with respect to the infection process, where
each node has an effective average degree $k=\kave$.

Different networks differ in the rate of rewiring that is required
for a change of the threshold. We have shown that for networks
with different degree distributions the relevant quantity is the
second moment of the degree distribution. Only if the rewiring is
larger than the second moment $\lc \gtrsim \ksqave$ then the
threshold is affected and is increased from
$\lc^{-1}=\kave/\ksqave$ to $\lc^{-1}=1/\kave$.

 For a finite system, even though a true threshold does not
exist, the crossover rewiring rate $\lcNL$ increases as we
increase the rewiring rate.  For homogeneous networks such as ER
networks rewiring has little effect on the behavior of the disease
since $\kave^2 \approx \ksqave$.  For heterogeneous networks such
as SF networks, the change is more significant. For SF networks
with $\gamma>3$ we have argued that in the thermodynamic limit the
threshold will increase continuously with the rewiring rate. On
the other hand, for SF networks with $\gamma<3$  there is no threshold
in the thermodynamic limit, except for an infinite
rewiring rate.

The support of the Israel Science Foundation (ISF) is gratefully
acknowledged. We thank Oren Shriki for discussions. YS thanks Aaron Clauset for computing resources.
A. R\'akos acknowledges financial support from the Hungarian Scientific Research
Fund (OTKA) grants PD-72604, PD-78433 and from the Bolyai Scholarship of
the Hungarian Academy of Sciences.

\bibliographystyle{apsrev}
\bibliography{net}

\end{document}